\documentclass[aps,prd,10pt,twocolumn,final,notitlepage,oneside,nobibnotes,
nofootinbib,superscriptaddress,showpacs,showkeys,centertags]{revtex4-1}

\usepackage{amsmath,amsfonts,amssymb}
\usepackage{bm}
\usepackage{dcolumn}
\usepackage[dvips]{graphicx,graphics,color,epsfig}
\usepackage{latexsym}
\usepackage{natbib}
\usepackage{verbatim}
\usepackage{bm}
\usepackage{tabulary}
\usepackage{natbib}

\begin{document}
\title{Oscillation characteristics of active and sterile neutrinos and
neutrino anomalies at short distances}
\author{\firstname{V.~V.}~\surname{Khruschov}}
\email{khruschov\_vv@nrcki.ru}
\affiliation{National Research Center Kurchatov Institute, Academician
Kurchatov Place 1, Moscow, 123182 Russia}
\affiliation{Center for Gravitation and Fundamental Metrology, VNIIMS,
Ozernaya Street 46, Moscow, 119361 Russia}
\author{\firstname{S.~V.}~\surname{Fomichev}}
\email{fomichev\_sv@nrcki.ru}
\affiliation{National Research Center Kurchatov Institute, Academician
Kurchatov Place 1, Moscow, 123182 Russia}
\affiliation{Moscow Institute of Physics and Technology (State University),
Institutskii Lane 9, Dolgoprudnyi, Moscow Region, 141700 Russia}
\author{\firstname{O.~A.}~\surname{Titov}}
\email{titov\_oa@nrcki.ru}
\affiliation{National Research Center Kurchatov Institute, Academician
Kurchatov Place 1, Moscow, 123182 Russia}

\begin{abstract}
A generalized phenomenological (3 + 2 + 1)-model with three active and three
sterile neutrinos is considered for the calculation of the neutrino oscillation
characteristics at normal mass hierarchy of active neutrinos and significant
splitting between the mass of one sterile neutrino and the masses of other two
sterile neutrinos. A new parametrization and a certain form of the general
neutrino mixing matrix for active and sterile neutrinos are suggested taking
into account the possible violation of {\it CP} invariance in the lepton
sector. The test values of the neutrino masses and the mixing parameters are
chosen. The transition probabilities for the different flavors of neutrino,
which can be transformed into each other, are calculated, and the graphical
dependences are obtained for the disappearance probability of muon
neutrino/antineutrino and appearance probability of electron
neutrino/antineutrino in the muon neutrino/antineutrino jet as a function of
distance and other model parameters at their acceptable values and at neutrino
energies not higher than $50$~MeV, as well as a function of the ratio of the
distance to the neutrino energy. It is shown that in the case of the mixing
matrix of a definite type (of the $a_2$ type) between active and sterile
neutrinos the explanation of the acceleration anomaly on small distances in the
neutrino data (LSND anomaly) is possible, as well as the reactor and gallium
anomalies. The theoretical results obtained can be used for the interpretation
and prediction of results of the ground-based neutrino experiments on searching
for the sterile neutrinos, and also for the analysis of certain astrophysical
observation data.
\end{abstract}



\maketitle

\section{\textnormal{INTRODUCTION}}
\label{Section1}
An one of the most important problems of the modern neutrino physics is the
problem of light sterile neutrinos, which is related to neutrino- and
antineutrino-flux anomalies observed at short distances in a number of
ground-based experiments \cite{Abazajian2012,Schwetz2011,Giunti2013,Kopp2013}.
The presence of such anomalies, if it will be confirmed at a sufficiently high
reliability level, is obviously beyond the Standard Model (SM), as well as the
Minimally Extended Standard Model (MESM) with three active neutrinos of
different mass, since oscillations of only three known active neutrinos cannot
explain these anomalies. Sterile neutrinos are new particles that do not
interact with the SM particles via the exchange by photons, $W$ and $Z$ bosons,
and naturally gluons \cite{Gorbunov2014}. The scale of sterile neutrino masses
that is required for interpretation of the anomalies mentioned above is about
$1$~eV.

In principle, sterile-neutrino mass values may belong to a broad energy range
from $10^{-5}$~eV to $10^{15}$~GeV \cite{deGouvea2005,Drewes2015}. It is
convenient to break down this range in such a way as to refer sterile neutrinos
of mass below $0.1$~eV to a class of ultralight sterile neutrinos, neutrinos of
mass between $0.1$ and $100$~eV to a class of light sterile neutrinos,
neutrinos of mass between $100$~eV and $10$~GeV to a class of heavy sterile
neutrinos, and neutrinos of mass above $10$~GeV to a class of superheavy
sterile neutrinos. Intensive experimental and theoretical studies are being
currently performed in order to solve the problem of light sterile neutrinos,
which are involved to interpret the accelerator anomaly observed in the
LNSD/MiniBooNe experiments \cite{Atha1996,Agu2001,Agu2013}, as well as the
reactor \cite{Mu2011,Hu2011} and gallium \cite{Abdu2009,Kae2010} anomalies, and
some astrophysical data \cite{Demi2014}. It is expected that experimental data
making it possible to confirm or disprove the existence of the aforementioned
anomalies will be obtained in the near future (see, for example,
Refs.~\cite{Abazajian2012,Gorbunov2014,Gav2011,Bel2013,Ser2015}).

Since the existence of sterile neutrinos is beyond the MESM framework,
different phenomenological models with one, two, or three sterile neutrinos
were proposed \cite%
{Abazajian2012,Schwetz2011,Giunti2013,Kopp2013,Gorbunov2014,Canetti2013,Conrad2013}.
In principle, the number of sterile neutrinos may be arbitrary as long as this
is compatible with the experimental data. If, however, to take into account the
possible existence of the left-right symmetry of weak interactions and to
associate sterile neutrinos with right-handed neutrinos, which are neutral with
respect to $SU(2)_L$-weak interactions, then the number of sterile neutrinos
should be equal to three, that is, it is necessary to consider a $(3+3)$-model
with three active and three sterile neutrinos \cite{Canetti2013,Conrad2013}. As
one such model, there is the $(3+1+2)$-model (or $(3+2+1)$-model) with three
sterile neutrinos \cite{Zysina2014,KhruFom2015}, where two of them are
approximately degenerate in mass, while the third one has a mass that may
differs markedly from the masses of the other two sterile neutrinos. The
$(3+2+1)$-model involving two ultralight sterile neutrinos and one light
sterile neutrino was used to evaluate the mass characteristics of both active
and sterile neutrinos \cite{Zysina2014} and to calculate the appearance and
survival probabilities for active and sterile neutrinos in the Sun with
allowance made for the coherent neutrino scattering in matter
\cite{KhruFom2015}. An enhancement of the yield of three sterile neutrinos in
high-density media as the neutron-to-proton number ratio approaches two was
considered in Ref.~\cite{PAZH2015}. This effect may have an influence on the
formation of neutrino fluxes in supernovae.

It is well known that there are difficulties in models with three active and
three sterile neutrinos in matching the whole number of neutrinos with data
coming from the cosmological observations (see, for example,
Refs.~\cite{Komatsy2011,Ade2013}). At the present time, the most popular models
are phenomenological models with one or two sterile neutrinos, that is the
so-called $(3+1)$- and $(3+2)$-models
\cite{Abazajian2012,Schwetz2011,Giunti2013,Kopp2013,Gorbunov2014}. The basic
argument for this choice is that, according to currently prevalent ideas, one
or two sterile neutrinos are sufficient for explaining the experimentally
observed anomalies. Moreover, the most recent data obtained on the basis of
standard cosmological models from observations of the cosmic microwave
background restrict the number of new relativistic neutrinos up to one
\cite{Komatsy2011,Ade2013}. Despite this, the present study is devoted to
considering a phenomenological model that involves three active and three
sterile neutrinos. There are several reasons for this choice. First, this is
the aforementioned principle of left-right symmetry, which is possibly restored
at a scale that exceeds the scale of spontaneous breaking of electroweak
symmetry (of the order of $1$~TeV). Second, the form chosen in
Sec.~\ref{Section3} for mixing matrix of both active and sterile neutrinos
dictates the introducing namely three sterile neutrinos, provided that the
mixing matrix satisfies the correct unitarity condition. And finally, one
cannot exclude that nonstandard cosmological models may prove to be viable, in
framework of which, despite the presence of three sterile neutrinos, their
effect is reduced or suppressed in data from cosmological observations (see,
for example, Refs.~\cite{HoScherrer2013,Chu2015}).

The paper is organized as follows. In Sec.~\ref{Section2}, we present the
experimental data on the oscillation characteristics of active neutrinos and
describe the experimental indications of possible existence of the anomalies
that go beyond the scope of the Minimally Extended Standard Model with three
active massive neutrinos. In Sec.~\ref{Section3}, the fundamentals of a
generalized neutrino model with three active and three sterile neutrinos are
discussed with allowing for the results reported in
Refs.~\cite{Zysina2014,KhruFom2015}. In Sec.~\ref{Section4}, we consider a way
to determine the oscillation properties of both active and sterile neutrinos on
the basis of equations that describe the propagation amplitudes for neutrino in
a vacuum. In the same section, we also present exact analytical expressions for
determining the appearance and disappearance (survival) probabilities for
neutrinos of specific flavor, which are particular cases of the general
expressions from Ref.~\cite{Bilenky} for the model under consideration. The
results of our numerical calculations of the oscillation properties of
neutrinos at test values of the neutrino masses and mixing parameters, which
are performed with allowance for the sterile-neutrino contributions, are given
in Sec.~\ref{Section5}. We obtain the appearance probabilities for electron
neutrinos/antineutrinos in beams of muon neutrinos/antineutrinos versus the
distance from the neutrino source and versus the ratio of this distance to the
neutrino energy, and also we obtain the disappearance probabilities for muon
neutrinos/antineutrinos at model-parameter values typical for the accelerator
anomaly. In the concluding Sec.~\ref{Section6}, we discuss the paper results,
which may be of use in interpreting and predicting the results of experiments
devoted to searching for the effects associated with sterile neutrinos, and
also in analyzing some astrophysical data.

\vspace{-5mm}
\section{\textnormal{OSCILLATION CHARACTERISTICS OF ACTIVE NEUTRINOS AND
ANOMALIES IN NEUTRINO DATA}}
\label{Section2}
As is well known, oscillations of solar, atmospheric, reactor, and accelerator
neutrinos may be explained by the mixing of different neutrino mass
eigenstates. This means that the flavor states of active neutrinos $(a)$ are a
mixture of at least three neutrino mass eigenstates, and vice versa. The mixing
of neutrino states is described in terms of the
Pontecorvo--Maki--Nakagawa--Sakata matrix $U_{PMNS}\equiv U\equiv VP$ as
\vspace{-1mm}
\begin{equation}
\psi^{a}_L=U^{a}_i\psi^i_L\,,
\vspace{-1mm}
\label{eq1}
\end{equation}
where $\psi^{a,i}_L$ are left-handed chiral fields of flavor or massive
neutrinos, respectively, and $a=\{e,\mu,\tau\}$, $i=\{1,2,3\}$. For three
active neutrinos, the matrix $V$ can be expressed in terms of a standard
parametrization \cite{Olive2014} as
\begin{widetext}
\begin{equation}
V=\left(\begin{array}{lcr} c_{12}c_{13} & s_{12}c_{13} & s_{13}e^{-i\delta}\\
-s_{12}c_{23}-c_{12}s_{23}s_{13}e^{i\delta} & c_{12}c_{23}-s_{12}s_{23}s_{13}
e^{i\delta} & s_{23}c_{13}\\
s_{12}s_{23}-c_{12}c_{23}s_{13}e^{i\delta} & -c_{12}s_{23}-s_{12}c_{23}s_{13}
e^{i\delta} & c_{23}c_{13}
\end{array}\right),
\label{eq2}
\end{equation}
\end{widetext}
where $c_{ij}\equiv\cos\theta_{ij}$, $s_{ij}\equiv\sin\theta_{ij}$,
$\delta\equiv\delta_{CP}$ is the phase associated with Dirac's {\it CP}
violation in the lepton sector, and $P={\rm diag}\{1,e^{i\alpha},e^{i\beta}\}$,
with $\alpha\equiv\alpha_{CP}$ and $\beta\equiv\beta_{CP}$ the phases
associated with Majorana's {\it CP} violation.

In general, a unitary $n\times n$ matrix is determined by $n^2$ real-valued
parameters, for which one can take $n(n-1)/2$ angles and $n(n+1)/2$ phases.
With allowance for the structure of the SM electroweak Lagrangian, which
includes currents composed of quark, charged-lepton, and neutrino fields, it is
possible to exclude $2n-1$ phases in the case of Dirac neutrino fields. But in
the case where the neutrino fields belong to the Majorana type, we can exclude
only $n$ phases associated with Dirac charged leptons. In view of this, a
$n\times n$ mixing matrix is determined either by $n(n-1)/2$ angles and
$(n-1)(n-2)/2$ phases if neutrinos are Dirac particles or by $n(n-1)/2$ angles
and $n(n-1)/2$ phases if neutrinos would be Majorana particles
\cite{Bilenky1977}. Thus, for three active Dirac neutrinos it is necessary to
determine three mixing angles $\theta_{12}$, $\theta_{13}$, $\theta_{23}$ and
one mixing phase ({\it CP} phase $\delta_{CP}$ ) in order to specify the mixing
matrix $U_{PMNS}$, while for three active Majorana neutrinos one needs to
specify the analogous three mixing angles ($\theta_{12}$, $\theta_{13}$,
$\theta_{23}$) and three {\it CP} phases $\delta_{CP}$, $\alpha_{CP}$, and
$\beta_{CP}$.

However, oscillation experiments with atmospheric, solar, reactor, or
accelerator neutrinos give no possibilities to measure the Majorana {\it CP}
phases and to attribute neutrinos to the Majorana or Dirac particle type.
Nevertheless, experimental results concerning neutrino oscillations suggest
violation of conservation laws of the lepton numbers $L_e$, $L_{\mu}$, and
$L_{\tau}$ and, because of nonzero values of two oscillation parameters
$\Delta m_{12}^2$ and $\Delta m_{13}^2$ (where $\Delta m_{ij}^2=m_i^2-m_j^2$),
the existence of at least two nonzero active-neutrino masses that are not equal
to each other.

Now we present the values for the neutrino mixing angles and mass-squared
differences with standard uncertainties at an $1\sigma$ level, which specify
three-flavor oscillations of light active neutrinos in a vacuum, which were
obtained from a global analysis of the most recent high-precision measurements
of neutrino oscillation parameters \cite{Capozzi}. Specifically, they are
\begin{subequations}
\begin{align}
&\sin^2\theta_{12}=0.308^{+0.017}_{-0.017}\,,\label{eq3a}\\
&\sin^2\theta_{23}=\left\{\begin{array}{lr}{\rm NH}:
& 0.437^{+0.033}_{-0.023}\\ {\rm IH}:
& 0.455^{+0.039}_{-0.031}\end{array}\right.\!\!,\label{eq3b}\\
&\sin^2\theta_{13}=\left\{\begin{array}{lr}{\rm NH}:
& 0.0234^{+0.0020}_{-0.0019}\\ {\rm IH}:
& 0.0240^{+0.0019}_{-0.0022}\end{array}\right.\!\!,\label{eq3c}\\
&\Delta m_{21}^2/10^{-5}{\rm eV}^2=7.54^{+0.26}_{-0.22}\,,\label{eq3d}\\
&\Delta m_{31}^2/10^{-3}{\rm eV}^2=\left\{\begin{array}{lr}{\rm NH}:
& 2.43^{+0.06}_{-0.06}\\ {\rm IH}:
& -2.38^{+0.06}_{-0.06}\end{array}\right.\!\!.\label{eq3e}
\end{align}
\label{eq3}
\end{subequations}\\
The {\it CP} phases $\alpha_{CP}$, $\beta_{CP}$, and $\delta_{CP}$ and the
neutrino mass scale are still not known at the present time. Since only the
absolute value of the oscillation characteristic
$\Delta m^2=m_3^2-(m_1^2+m_2^2)/2$ is known, the absolute values of the
neutrino masses can be ordered by two ways, namely, $a)\; m_1<m_2<m_3$ and
$b)\; m_3<m_1<m_2$, that is, it may be either a normal hierarchy [NH, case (a)]
or an inverse hierarchy [IH, case (b)] in the neutrino-mass spectrum.

Along with the values presented in Eqs.~(\ref{eq3}) for the oscillation
characteristics of the neutrinos, indications that the neutrino fluxes data for
some processes have anomalies that cannot be explained by oscillations of only
active (that is, electron, muon, and tau) neutrinos and antineutrinos have
existed for a rather long time. These anomalies include the LSND (accelerator)
anomaly \cite{Atha1996,Agu2001,Agu2013} and the reactor \cite{Mu2011,Hu2011}
and gallium (calibration) \cite{Abdu2009,Kae2010} anomalies, which can be
explained by the existence of one or two extra neutrinos that do not interact
with other SM particles, that is, sterile neutrinos. The characteristic scale
of sterile-neutrino masses that is required for describing the aforementioned
anomalies is about $1$~eV. These anomalies manifest themselves at short
distances (more precisely, at distances $L$ such that the numerical values of
the parameter $\Delta m^2 L/E$, where $E$ is the neutrino energy, are about
unity). The experiments at such distances are referred to as short-baseline
(SBL) experiments. The possible excess of the electron antineutrino fraction in
beams of muon antineutrinos in relation to what is expected within the MESM
framework is referred to as the LSND anomaly (sometimes accelerator anomaly).
Similar results were also observed in the MiniBooNE experiments for electron
neutrinos and antineutrinos (see, however, Ref.~\cite{Giunti2013}). A deficit
of reactor electron antineutrinos at short distances was called as the reactor
anomaly, while a deficit of electron neutrinos from a radioactive source in the
calibration of the detectors for the SAGE and GALLEX experiments is usually
referred to as the gallium (calibration) anomaly. In other words, data on the
anomalies concern both the appearance of electron neutrinos or antineutrinos in
the fluxes of muon neutrinos or antineutrinos, respectively, and the
disappearance of electron or muon neutrinos or antineutrinos. It should be
emphasized that these three types of anomalies observed to date in the neutrino
fluxes data suggest the existence of light sterile neutrinos.

It should be noted that recent data coming from astrophysical observations and
concerning the formation of galaxies and their clusters can be explained by the
existence of light or heavy sterile neutrinos. In astrophysics, such sterile
neutrinos are real candidates for dark-matter particles
\cite{Dodelson1994,Kusenko2009,Demi2014}. At the present time, multicomponent
dark-matter models are frequently used to describe dark-matter-dominated
objects such as some galaxies and galaxy clusters \cite{Demi2014}. More details
on the possible existence of sterile neutrinos, their properties, and their
relation to still unknown dark-matter particles can be found in numerous papers
(see, for example, Refs.~\cite{Kusenko2009,Abazajian2012,Demi2014,Liao,An}).

\vspace{-5mm}
\section{\textnormal{FUNDAMENTALS OF THE PHENOMENOLOGICAL (3 + 2 + 1)-MODEL OF
ACTIVE AND STERILE NEUTRINOS}}
\label{Section3}
Below, we consider a phenomenological $(3+2+1)$ neutrino model, which is used
to include three sterile neutrinos in the formalism of theory of weak
interactions. This model can be employed both to obtain, with allowance made
for sterile neutrinos, the properties of neutrino fluxes studied in
ground-based SBL experiments and to analyze some astrophysical data, including
the data on neutrino radiation from supernovae
\cite{Zysina2014,KhruFom2015,PAZH2015}. The model under consideration includes
three active neutrinos and three sterile neutrinos, such that two sterile
neutrinos are approximately degenerate in mass, while the mass of the third
sterile neutrino may differ significantly from the masses of the other two
sterile ones.

Taking into account the results obtained in Refs.~\cite{Zysina2014,KhruFom2015},
let us consider the neutrino-mixing formalism for the case of three active and
three sterile neutrinos. We will use the indices $x$, $y$, and $z$ to
distinguish between the different sterile-neutrino types (flavors) and the
indices $1'$, $2'$, and $3'$ to distinguish between the extra massive states.
Further, we will denote by $s$ the set of indices $x$, $y$, and $z$ and by $i'$
the set of indices $1'$, $2'$, and $3'$. A general $6\times 6$ mixing matrix
$\widetilde{U}$ can then be expressed in terms of the $3\times 3$ matrices $S$,
$T$, $V$, and $W$ as
\begin{equation}
\left(\begin{array}{c}
\nu_a\\ \nu_s \end{array}\right)=
\left(\begin{array}{cc}
S&T\\ V&W\end{array}\right)
\left(\begin{array}{c}
\nu_i\\ \nu_{i'}\end{array}\right).
\label{eq4}
\end{equation}
The neutrino masses will be specified by means of the set
$\{m\}=\{m_i,m_{i'}\}$ ordered normally for $\{m_i\}$ as $\{m_1,m_2,m_3\}$ and
inversely for $\{m_{i'}\}$ as $\{m_{3'},m_{2'},m_{1'}\}$. For the unitary
$6\times 6$ matrix $\widetilde{U}$, we will consider only some particular cases
rather than the most general form, involving additional physical assumptions.
On the basis of data available from astrophysical and laboratory observations,
we assume that the mixing between active and sterile neutrinos is small. In
addition, the basis for sterile-neutrino mass eigenstates will be constructed
from states for which the matrix $W$ is approximately equal to the identity
matrix. In doing this, we restrict ourselves to a diagonal matrix of the form
$W=\widetilde\varkappa I$, where $I$ is the identity matrix and
$\widetilde\varkappa$ is a complex-valued parameter whose modulus is close to
unity, and for which we employ the form
$\widetilde\varkappa=\varkappa\exp{(i\phi)}$. The matrix $S$ will be
represented in the form
\begin{equation}
S=U_{PMNS}+\Delta U_{PMNS},
\label{eq5}
\end{equation}
where the matrix $\Delta U_{PMNS}$, as well as the matrix $T$ in
Eq.~(\ref{eq4}) should be small towards $U_{PMNS}$. In order to the
quantitative estimations of both the mixing between active and sterile
neutrinos and the sterile-neutrino-induced corrections to mixing between active
neutrinos to be convenient, we assume that
\begin{equation}
\Delta U_{PMNS} = -\epsilon U_{PMNS},
\label{eq6}
\end{equation}
where $\epsilon$ is a small quantity of the form $\epsilon=1-\varkappa$. The
matrix $S$ then has the form $S=\varkappa U_{PMNS}$, where $U_{PMNS}$ is the
known unitary $3\times 3$ matrix of active-neutrino mixing
($U_{PMNS}U_{PMNS}^+=I$).

Thus, upon the respective normalization that was chosen, the active neutrinos
are mixed, as it should be in the MESM, according to the
Pontecorvo--Maki--Nakagawa--Sakata matrix. Taking into account that the mixing
between active and sterile neutrinos is small and is determined by the small
parameter $\epsilon$, we choose the matrix $T$ as $T=\sqrt{1-\varkappa^2}\,a$,
where $a$ is an arbitrary unitary $3\times 3$ matrix, so that $aa^+=I$. The
$6\times 6$ mixing matrix $\widetilde{U}$, which can be written now in the form
of
\begin{align}
&\widetilde{U}=\left(\begin{array}{cc}
\varkappa U_{PMNS}&T\\
-e^{i\phi}T^+U_{PMNS}&\varkappa e^{i\phi}I
\end{array}\right) \nonumber \\
&\equiv
\left(\begin{array}{cc}
\varkappa U_{PMNS}&\sqrt{1-\varkappa^2}\,a\\
-e^{i\phi}\sqrt{1-\varkappa^2}\,a^+U_{PMNS}&\varkappa e^{i\phi}I
\end{array}\right),
\label{eq7}
\end{align}
is then strictly unitary. In Refs.~\cite{Zysina2014,KhruFom2015}, use was made
of an approximately unitary mixing matrix of the form analogous to
Eq.~(\ref{eq7}) at $\phi=0$, where, however, the condition requiring the
conservation of normalization for the sterile-neutrino states was
not taken into account (corrections associated with a nonunitarity of the
mixing matrix were discussed, for example, in Ref.~\cite{Antu2014})). In the
present study, we take $\phi=\pi/4$ for the test value of the unknown phase
$\phi$.

Although the value of the {\it CP} phase $\delta_{CP}$ has not yet been
established experimentally, estimates for it were obtained in a number of
studies (see, for example, Refs.~\cite{Capozzi,Khru2013,Petkov}) for the case
of a normal hierarchy for the active-neutrino mass spectrum, namely,
$\sin\delta_{CP} < 0$ and $\delta_{CP}\approx -\pi/2$. A normal hierarchy also
becomes more preferable upon taking into account the constraints on the sum of
the neutrino masses from data based on the cosmological observations
\cite{Huang}. In our further numerical calculations, we therefore restrict
ourselves to the NH case, setting $\delta_{CP}=-\pi/2$.

In the present study, we consider three particular cases for the form of the
matrix $a$, which specifies mixing between active and sterile neutrinos. They
will be denoted by $a_1$, $a_2$, and $a_3$. For the first two cases, we
choose these matrices in the form
\begin{equation}
a_{1}=\left(\begin{array}{lcr}
e^{-i\chi_1} & 0 & 0\\
0 & \cos\eta_1 & \sin\eta_1\\
0 & -\sin\eta_1 & \cos\eta_1\end{array}\right),
\label{eq8}
\end{equation}
\vspace{-3.5mm}
\begin{equation}
a_{2}=\left(\begin{array}{lcr}
\cos\eta_2 & \sin\eta_2 & 0\\
-\sin\eta_2 & \cos\eta_2 & 0\\
0 & 0 & e^{-i\chi_2}\end{array}\right),
\label{eq9}
\end{equation}
where $\chi_1$ and $\chi_2$ are the mixing phases for active and sterile
neutrinos, while $\eta_1$ and $\eta_2$ are their mixing angles. In our
calculations, we will use the following test values for the new mixing
parameters:
\begin{equation}
\chi_1=\chi_2=-\pi/2, \quad \eta_1= 5^{\circ}, \quad
\eta_2=\pm 30^{\circ}.
\label{eq10}
\end{equation}
In the third case, we specify the matrix $a_3$ in a diagonal form with two
phases $\zeta_1=\pi/2$ and $\zeta_2=\pi/4$, so that
\begin{equation}
a_3={\rm diag}\{e^{-i\zeta_1},e^{-i\zeta_2},1\}.
\label{eq11}
\end{equation}
In all three cases, we assume that the small parameter $\epsilon$ satisfies the
condition $\epsilon\lesssim 0.03$.

For the active neutrinos, we will use the results obtained in
Refs.~\cite{Zysina2014,KhruFom2015} on estimating the absolute values of the
masses $m_i$ ($i=1,2,3$) of light active neutrinos for the NH case (in eV
units):
\begin{equation}
m_1\approx 0.0016, \quad m_2\approx 0.0088, \quad m_3\approx 0.0497\,.
\label{eq12}
\end{equation}
We can now examine three versions of the generalized $(3+3)$-model that
correspond to the so-called $(3+1)$- and $(3+2)$-models of active and sterile
neutrinos. As was indicated above, these models are used to explain the
anomalies that manifest themselves in neutrino data at short distances.

In the first version of the model, which corresponds to the $(3+1)$-model, we
choose a value for the sterile-neutrino mass $m_{3'}$ in the region around
$1$~eV in accordance with the results obtained in Ref.~\cite{Kopp2013} for the
best value of the mass of only one sterile neutrino in this region:
$m_{3'}\approx 0.96$~eV. We specify the masses $m_{2'}$ and $m_{1'}$ of the
other two sterile neutrinos in the region around $30$~eV by using the result
obtained in Ref.~\cite{Demi2014} by fitting the mass of the warm-dark-matter
particle. We assume that the particles of two components of warm dark matter
have masses belonging to this range and taking values of $m_{2'}\approx 28$~eV
and $m_{1'}\approx 32$~eV. Within this version, we take the matrix $a_1$ for
the matrix $a$.

In the second version of our model, which corresponds to the $(3+2)$-model, we
choose values for the masses $m_{3'}$ and $m_{2'}$ of two sterile neutrinos in
the region around $1$~eV, following the result reported in Ref.~\cite{Kopp2013}
for the best values of the sterile-neutrino masses in the $(3+2)$-model, that
is, $m_{3'}\approx 0.69$~eV and $m_{2'}\approx 0.93$~eV. For the mass $m_{1'}$
of the third sterile neutrino, we make use of the mass of the possible
dark-matter particle \cite{Bulb2014,Boya2014,Horiuchi}, that is,
$m_{1'}\approx 7100$~eV; on the other hand, the choice of this value has
virtually no effect on the results obtained below. In this version, we take the
matrix $a_2$ for the matrix $a$.

In the third version, which we will call the $(3+1+1+1)$-model, we choose
values for the masses $m_{3'}$ and $m_{2'}$ of two sterile neutrinos in the
region around $1$~eV according to the results obtained in Ref.~\cite{Kopp2013}
for the smallest and the largest values of the possible sterile-neutrino masses,
that is, $m_{3'}\approx 0.32$~eV and $m_{2'}\approx 3.15$~eV. We recall that
these masses correspond to the boundaries of the interval of values admissible
for the squares of sterile-neutrino masses, namely,
$0.1$~eV$^2<m_s^2<10$~eV$^2$. The mass of the third sterile neutrino, $m_{1'}$,
is determined by the result obtained in Ref.~\cite{Demi2014} by fitting the
mass of the warm-dark-matter particle, that is, $m_{1'}\approx 30$~eV. In this
version, we take the matrix $a_3$ for the matrix $a$.

Thus, the absolute values of the masses of all neutrinos for three versions of
the model (in eV units) can be represented in the form
\begin{subequations}
\begin{align}
&m_{\nu,1}=\{0.0016,0.0088,0.0497,0.96,28,32\},\label{eq13a}\\
&m_{\nu,2}=\{0.0016,0.0088,0.0497,0.69,0.93,7100\},\label{eq13b}\\
&m_{\nu,3}=\{0.0016,0.0088,0.0497,0.32,3.15,30\}.\label{eq13c}
\end{align}
\label{eq13}
\end{subequations}
Taking into account these mass distributions of sterile neutrinos, we can call
the above three versions of the $(3+3)$ neutrino model, respectively, as the
$(3+1+2)$-model, $(3+2+1)$-model, and $(3+1+1+1)$-model of active and sterile
neutrinos. Note that these models can in principle mimic the $(3+1)$- and
$(3+2)$-models if the contribution of one or two sterile neutrinos decreases,
for example, upon a significant increase in their masses or upon a special
choice of the mixing angles.

\vspace{-5mm}
\section{\textnormal{EQUATIONS FOR NEUTRINO PROBABILITY AMPLITUDES AND
PROBABILITIES OF OSCILLATION TRANSITIONS BETWEEN ACTIVE AND STERILE NEUTRINOS}}
\label{Section4}
The equations for the flavor amplitudes of neutrino propagation in a vacuum for
the case of three active neutrinos are well known (see, e.g.,
Ref.~\cite{PAZH2015}) and are given by
\begin{equation}
i\partial_{r}\left(\begin{array}{c}
a_{e}\\ a_{\mu}\\ a_{\tau}
\end{array}\right)=H
\left(\begin{array}{c}
a_{e}\\ a_{\mu}\\ a_{\tau}\end{array}\right),
\label{eq14}
\end{equation}
where the matrix $H$ is expressed in terms of the
Pontecorvo--Maki--Nakagawa--Sakata mixing matrix $U_{PMNS}\equiv U$ as
\begin{equation}
H=\frac{U}{2E}\!\!\left(\!\!\begin{array}{ccc}
m_{1}^{2}-m_{0}^{2} & 0 & 0 \\
0 & m_{2}^{2}-m_{0}^{2} & 0 \\
0 & 0 & m_{3}^{2}-m_{0}^{2}
\end{array}\!\!\right)\!\!U^{+}.
\label{eq15}
\end{equation}
Here, $m_0$ is the lowest neutrino mass value among the three masses $m_1$,
$m_2$, and $m_3$, while $E$ is the neutrino energy.

In the general case when the number of different neutrino flavors is $3+N$, the
matrix $\Delta_{m^2}$ of the mass differences can be defined as
\begin{equation}
\Delta_{m^2}={\rm diag}\{m_1^2-m_0^2,m_2^2-m_0^2,\ldots,m_{3+N}^2-m_0^2\}\,,
\label{eq16}
\end{equation}
where $m_i$ ($i=1,2,\ldots 3+N$) are the neutrino masses and $m_0$ is the
lowest neutrino mass among all $m_i$. In the case of the $(3+3)$-model, we
obtain the following equations for the neutrino propagation, which generalize
Eqs.~(\ref{eq14}) and (\ref{eq15}) and have the form
\begin{equation}
i\partial_{r}\left(\begin{array}{c}
a_{a}\\ a_{s}
\end{array}\right)=\frac{\widetilde{U}}{2E}\Delta_{m^2}\widetilde{U}^+
\left(\begin{array}{c}
a_{a}\\ a_{s}
\end{array}\right),
\label{eq17}
\end{equation}
where $\widetilde{U}$ is the unitary $6\times 6$ mixing matrix for active and
sterile neutrinos that is specified by Eq.~(\ref{eq7}). For antineutrinos, the
respective equations have the form
\begin{equation}
i\partial_{r}\left(\begin{array}{c}
{a}_{\overline{a}}\\ {a}_{\overline{s}}
\end{array}\right)=\frac{\widetilde{U}^{*}}{2E}\Delta_{m^2}\widetilde{U}^T
\left(\begin{array}{c}
{a}_{\overline{a}}\\ {a}_{\overline{s}}
\end{array}\right),
\label{eq18}
\end{equation}
where $\ast$ denotes complex conjugation. While solving these equations with
taking into account the oscillation parameter values given in
Sec.~\ref{Section3}, we can find the survival probabilities for electron or
muon neutrinos or antineutrinos, as well as the appearance or disappearance
probabilities for neutrinos or antineutrinos of any other flavor, versus the
neutrino/antineutrino energy and versus the distance from the source.

From Eqs.~(\ref{eq17}) and (\ref{eq18}) one can obtain analytical expressions
for transitions between neutrino/antineutrino flavors in a vacuum versus the
distance $L$ from the source. If $\widetilde{U}$ is the generalized $6\times 6$
mixing matrix of the form given by Eq.~(\ref{eq7}) and if we introduce the
notation $\Delta_{ki}\equiv\Delta m_{ik}^2L/(4E)$, then, according to
Ref.~\cite{Bilenky}, one can calculate the probabilities for the transitions of
$\nu_{\alpha}$ to $\nu_{\alpha^{\prime}}$ or of $\overline{\nu}_{\alpha}$ to
$\overline{\nu}_{\alpha^{\prime}}$ by the formula
\begin{align}
&P(\nu_{\alpha}(\overline{\nu}_{\alpha})\rightarrow\nu_{\alpha^{\prime}}
(\overline{\nu}_{\alpha^{\prime}}))=\delta_{\alpha^{\prime}\alpha}-\nonumber\\
&-4\sum_{i>k}{\rm Re}(\widetilde{U}_{\alpha^{\prime} i}
\widetilde{U}_{\alpha i}^{\ast}\widetilde{U}_{\alpha^{\prime} k}^{\ast}
\widetilde{U}_{\alpha k})\sin^2\Delta_{ki}\,\pm\nonumber \\
&\pm2\sum_{i>k}{\rm Im}(\widetilde{U}_{\alpha^{\prime} i}
\widetilde{U}_{\alpha i}^{\ast}\widetilde{U}_{\alpha^{\prime} k}^{\ast}
\widetilde{U}_{\alpha k})\sin 2\Delta_{ki}\,,
\label{eq19}
\end{align}
where the upper sign $(+)$ corresponds to the
$\nu_{\alpha}\rightarrow\nu_{\alpha^{\prime}}$ neutrino transitions, while the
lower sign $(-)$ corresponds to the
$\overline{\nu}_{\alpha}\rightarrow\overline{\nu}_{\alpha^{\prime}}$
antineutrino transitions. Note that the flavor indices $\alpha$ and
$\alpha^{\prime}$ (just as the summation indices $i$ and $k$ for massive
states) refer to all neutrinos, both active and sterile ones. Moreover, it
follows from Eq.~(\ref{eq19}) that the relation
$P(\nu_{\alpha}\rightarrow\nu_{\alpha})\equiv P(\overline{\nu}_{\alpha}
\rightarrow\overline{\nu}_{\alpha})$ holds exactly as a consequence of
{\it CPT} conservation \cite{Bilenky}. In order to control the accuracy of
numerical results obtained on the basis of Eqs.~(\ref{eq17}) and (\ref{eq18})
with allowance made for the fine sterile-neutrino effects, the calculations
were also performed with using the exact analytical expressions (\ref{eq19}).

\vspace{-5mm}
\section{\textnormal{NUMERICAL RESULTS FOR OSCILLATIONS OF ACTIVE AND STERILE
NEUTRINOS}}
\label{Section5}
In the present study, we focus primarily on the possibility of describing,
within the model versions being considered, the anomalies that were found in
data from the LSND experiment on the oscillations of accelerator muon neutrinos
and antineutrinos, which afterwards were tested and will be still tested in a
number of future accelerator experiments \cite{Gariazzo}. This refers to data
on the disappearance of muon neutrinos and antineutrinos and to the appearance
of electron neutrinos and antineutrinos in the processes
$\overline{\nu}_{\mu}\to\overline{\nu}_{e}$ and ${\nu}_{\mu}\to{\nu}_{e}$. A
typical ratio of the distance travelled by a neutrino before detection to the
neutrino energy is either several meters per megaelectronvolt or one meter per
several megaelectronvolts. Attempts at a simultaneous description of all data
in these processes lead to difficulties. In particular, problems associated
with different values of the excess of the ${\nu}_{e}$ and $\overline{\nu}_{e}$
yields in the MiniBooNE experiment can be solved upon admitting {\it CP}
violation \cite{Mal2007,Pal2005,Kar2007}. It should be noted that the reactor
and gallium anomalies manifesting themselves in neutrino data as the
disappearance of electron neutrinos and antineutrinos can be described within
the model being considered by appropriate choosing a value of the parameter
$\epsilon$. This leads to values smaller than unity for the parameter
$\varkappa=1-\epsilon$ appearing in expression (\ref{eq7}) and, hence, to a
deficit of electron neutrinos and antineutrinos.

\begin{figure*}
\includegraphics[width=0.875\textwidth]{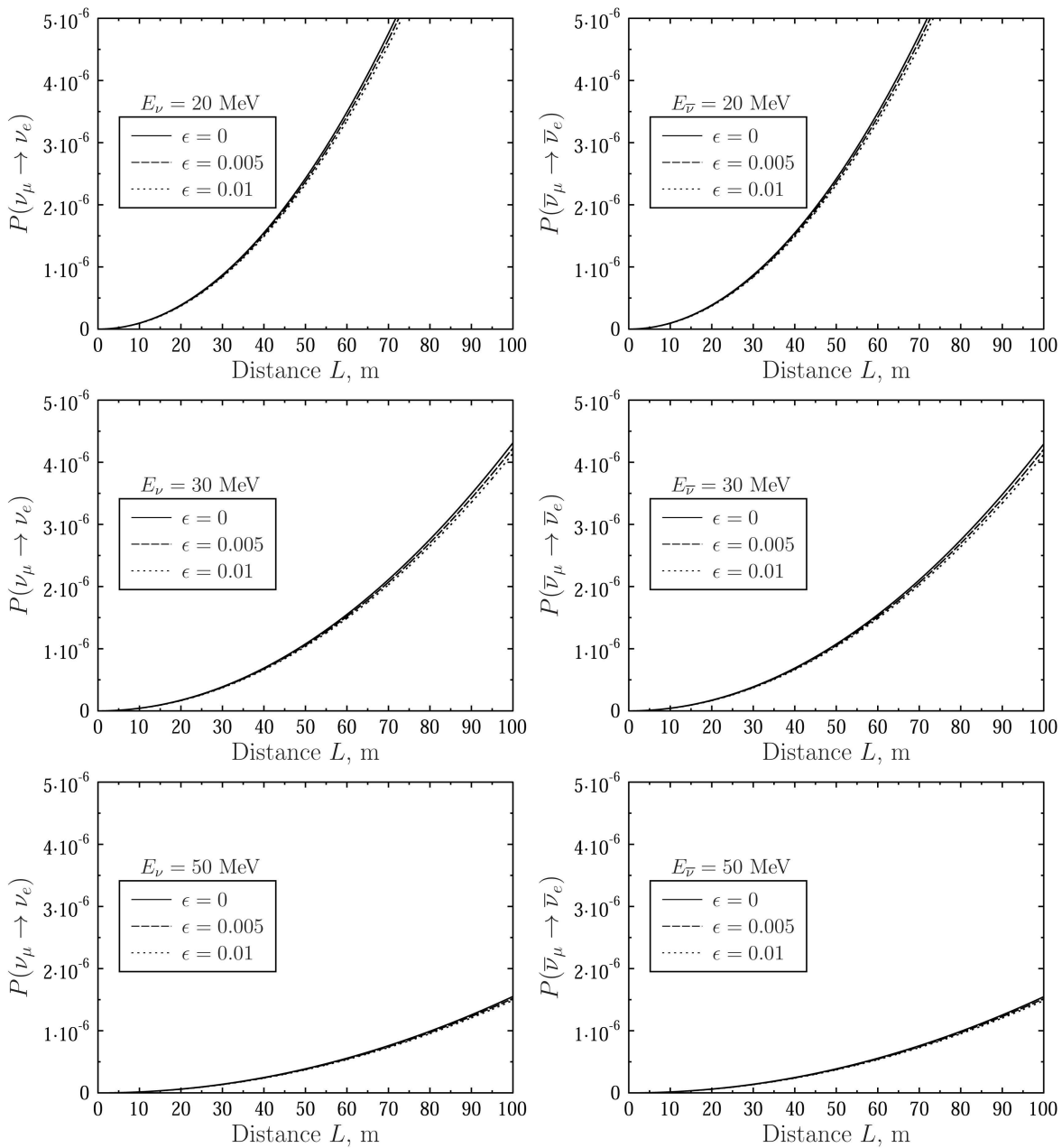}
\caption{Appearance probability for electron (left-hand panels) neutrinos and
(right-hand panels) antineutrinos as a function of the distance from the source
in beams of muon neutrinos and antineutrinos, respectively, at various
neutrino-beam energies and various values of coupling constant $\epsilon$ of
active and sterile neutrinos for two mixing matrices $a_1$ and $a_3$ leading to
identical results.}
\label{Fig1}
\end{figure*}
\begin{figure*}
\includegraphics[width=0.87\textwidth]{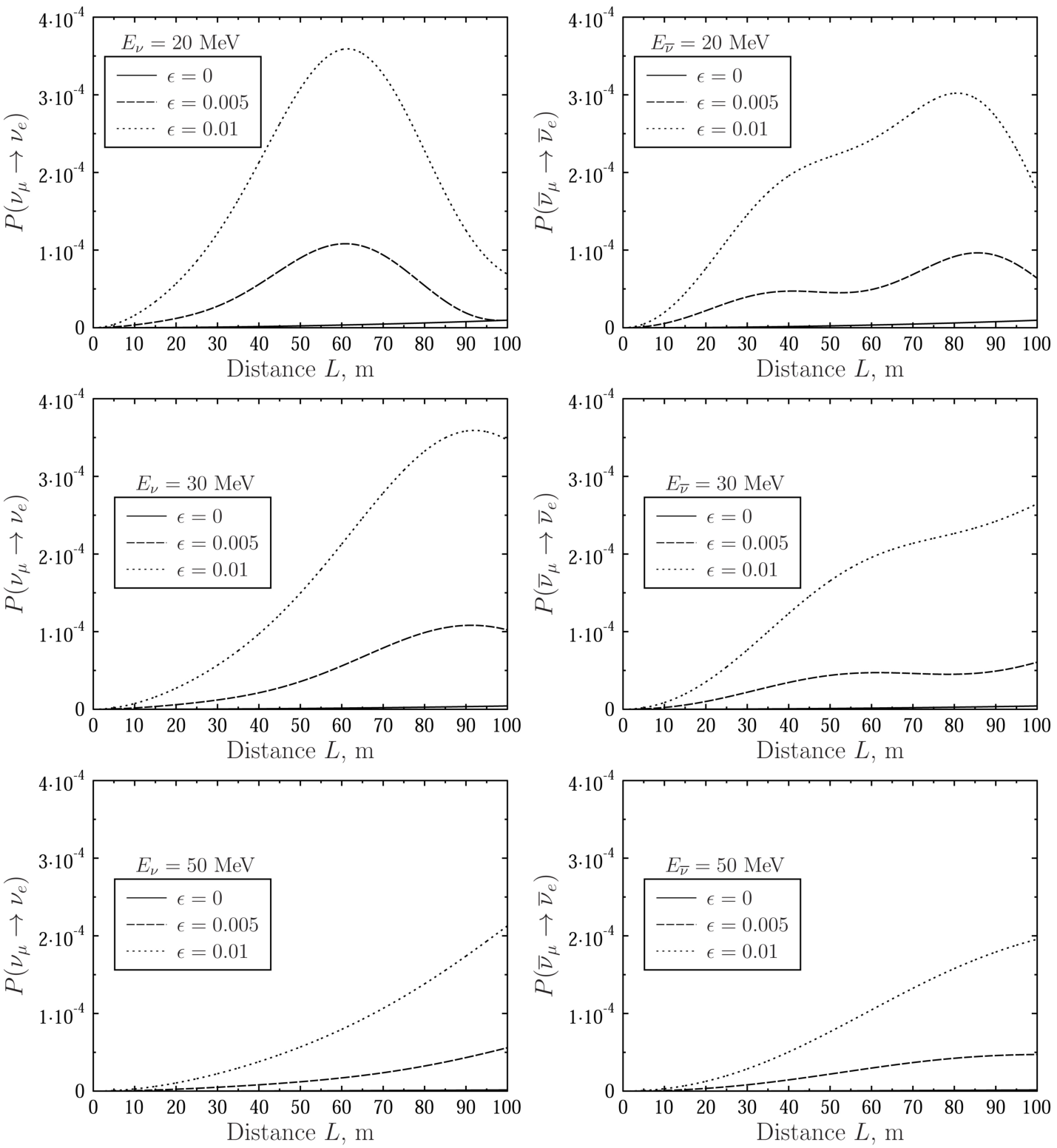}
\caption{Appearance probability for electron (left-hand panels) neutrinos and
(right-hand panels) antineutrinos as a function of the distance from the source
in beams of muon neutrinos and antineutrinos, respectively, at various
neutrino-beam energies and various values of coupling constant $\epsilon$ of
active and sterile neutrinos for the case of mixing matrix $a_2$ at
$\eta_2=\pi/6$.}
\label{Fig2}
\end{figure*}

The appearance probabilities for electron neutrinos and antineutrinos in
accelerator beams of muon neutrinos and antineutrinos are shown in
Figs.~\ref{Fig1} and \ref{Fig2}, respectively, versus the distance from the
source (from zero up to $100$~m) at various values of the neutrino energy in
the range between $20$ and $50$~MeV. Shown in the same figures for the sake of
comparison are the dependences obtained for the analogous probabilities within
the model containing only three active neutrinos, that is, in the absence of
sterile-neutrino contributions (at $\epsilon=0$).

In Fig.~\ref{Fig1}, the dependences of the appearance probabilities for
electron (left-hand panels) neutrinos and (right-hand panels) antineutrinos in
beams of muon neutrinos and antineutrinos, respectively, on the distance from
the source are shown at various neutrino-beam energies and at various values of
coupling constant $\epsilon$ of active and sterile neutrinos for two cases of
mixing matrices, $a_1$ and $a_3$. In these two cases, the results are found to
be identical, and, as can be seen from Fig.~\ref{Fig1}, the effect of sterile
neutrinos on the appearance probability for electron neutrinos/antineutrinos in
these cases is negligibly small in the range of values up to $\epsilon=0.01$ of
the coupling constant of active and sterile neutrinos. At the $\epsilon$-values
considered here, the neutrino and antineutrino yields are nearly identical at
neutrino energies of interest between $20$ and $50$~MeV. According to the
results of our calculations, the distinction between the yields of electron
neutrinos and antineutrinos for these mixing-matrix versions that is associated
with a nonzero value of the phase $\delta_{CP}$ manifests itself at the
distances under consideration only at neutrino energies not higher than $1$~MeV
and is weakly dependent on the parameter $\epsilon$.

In Fig.~\ref{Fig2}, the dependences of the appearance probabilities for
electron (left-hand panels) neutrinos and (right-hand panels) antineutrinos in
beams of muon neutrinos and antineutrinos, respectively, on the distance from
the source are shown at various neutrino-beam energies and various values of
coupling constant $\epsilon$ of active and sterile neutrinos for the mixing
matrix $a_2$ at the test parameter value of $\eta_2=\pi/6$. One can see that
this case differs drastically from the previous cases where the mixing matrix
is $a_1$ or $a_3$, the results for which are presented in Fig.~\ref{Fig1} at the
neutrino energies between $20$ and $50$~MeV. Firstly, the relative effect of
sterile neutrinos is by and large substantially stronger in the current case and
has the character of oscillations, as can be seen from the results for the
neutrino energy of $20$~MeV. Moreover, the relative neutrino and antineutrino
yields at the parameter value of $\epsilon=0.005$, for example, are
approximately two orders of magnitude greater (about $\sim 10^{-4}$) than
those at $\epsilon=0$ (about $\sim 10^{-6}$). Secondly, the yields of electron
neutrinos and antineutrinos are markedly different at this value of the
parameter $\epsilon$ in the neutrino-energy range of $20\div50$~MeV that we
consider. The appearance probabilities for electron antineutrinos and neutrinos
are different owing to {\it CP} violation, but, without sterile neutrinos (that
is, at $\epsilon=0$), this difference is extremely small at neutrino energies in
the range of $20\div50$~MeV. However, for the case of mixing matrix of the
$a_2$-type it becomes quite distinct upon taking into account sterile neutrinos
at the above values of the parameter $\epsilon$.

\begin{figure*}
\includegraphics[width=0.89\textwidth]{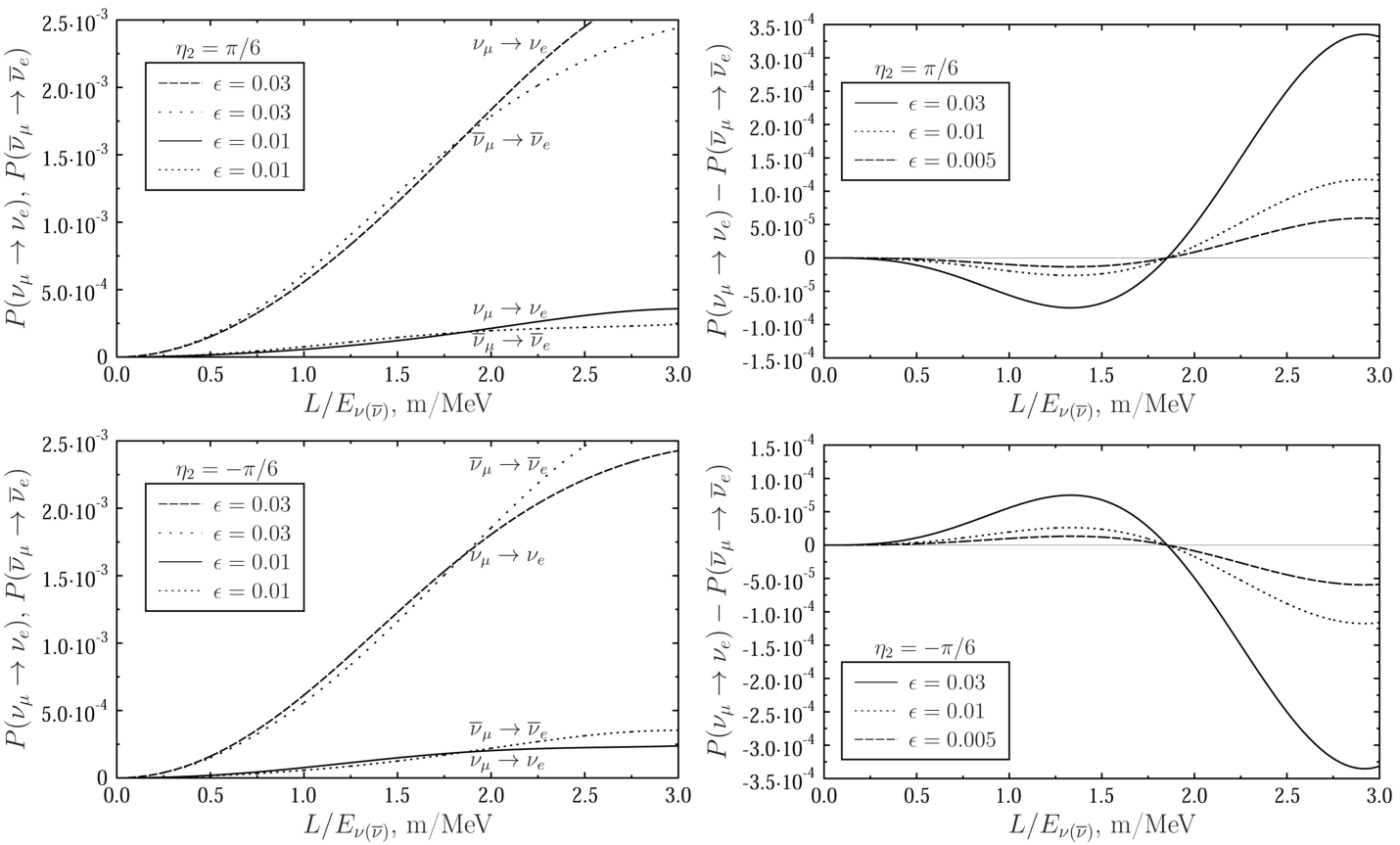}
\caption{Appearance probabilities for electron neutrinos and antineutrinos in
beams of muon neutrinos and antineutrinos, respectively, versus the ratio of
the distance from the source to the neutrino energy at various values of
coupling constant $\epsilon$ of active and sterile neutrino for the case of
mixing matrix $a_2$ (left-hand panels) and difference of these probabilities
(right-hand panels) at $\eta_2=\pi/6$ (upper panels) and $\eta_2=-\pi/6$
(lower panels).}
\label{Fig3}
\end{figure*}
\begin{figure*}
\includegraphics[width=0.87\textwidth]{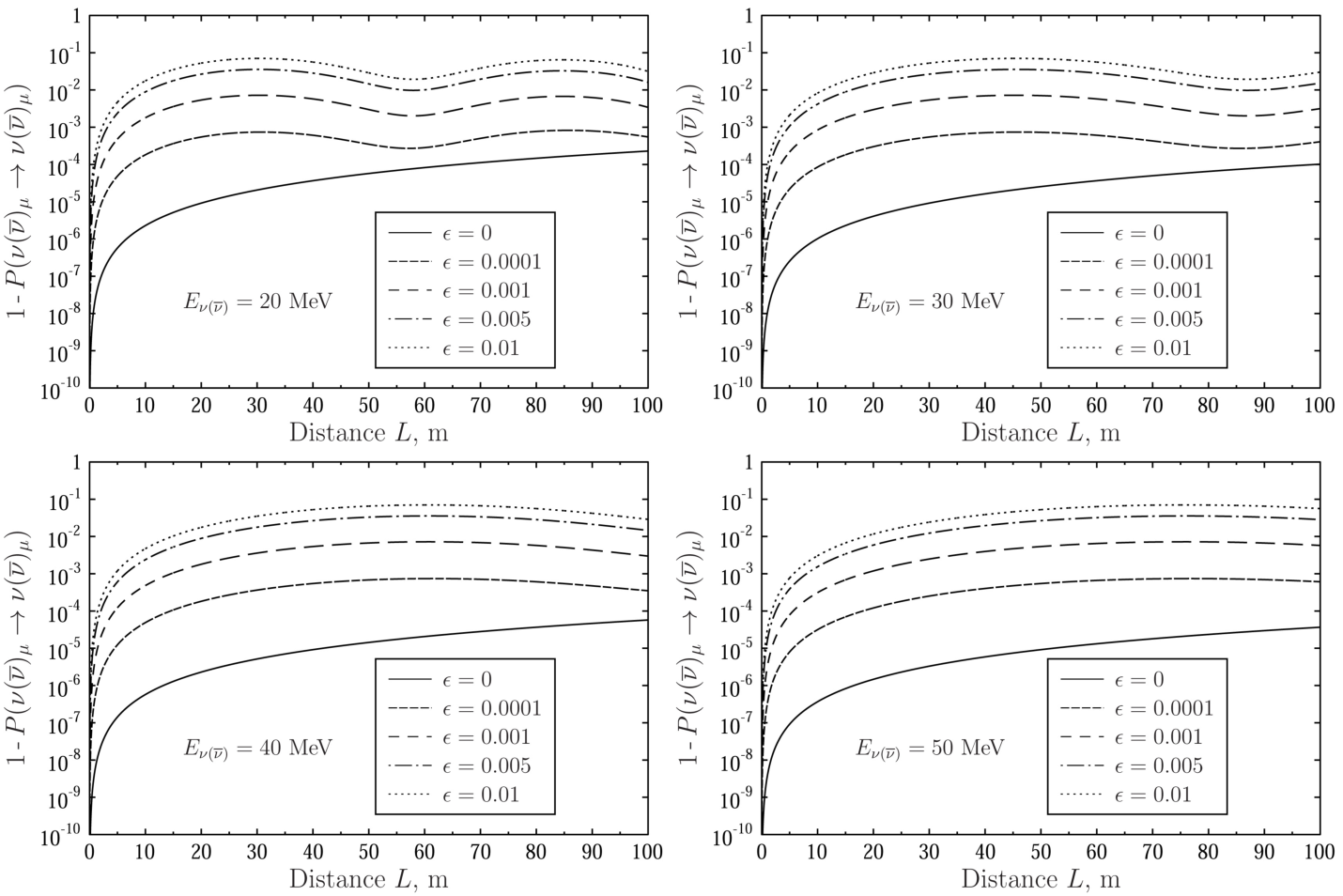}
\caption{Disappearance probability for muon neutrinos (antineutrinos) as a
function of the distance from the source in beams of muon neutrinos
(antineutrinos) at various energies of neutrino beams and various values of
coupling constant $\epsilon$ of active and sterile neutrinos for the case of
mixing matrix $a_2$.}
\label{Fig4}
\end{figure*}

\begin{figure*}
\includegraphics[width=0.85\textwidth]{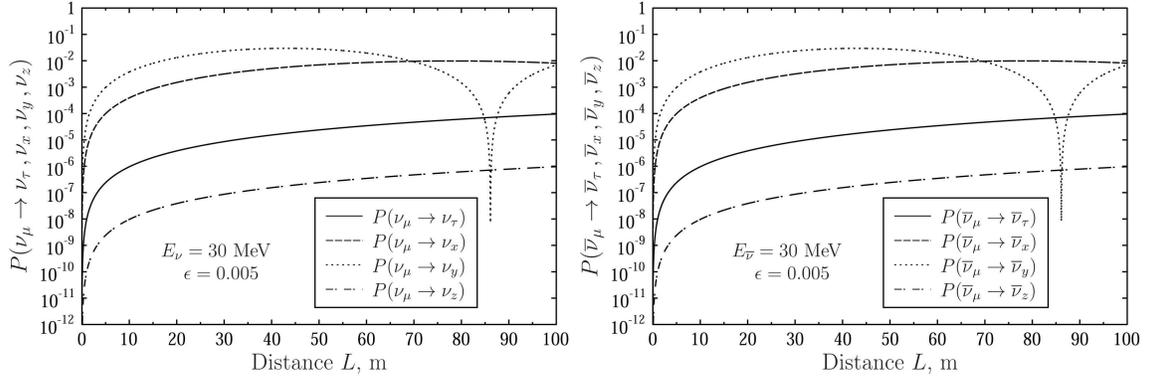}
\caption{Appearance probabilities for tau and three sterile (left-hand panel)
neutrinos and (right-hand panel) antineutrinos versus the distance from the
source in beams of muon neutrinos and antineutrinos, respectively, at the
neutrino-beam energy $30$~MeV and at the value of $\epsilon=0.005$ of coupling
constant between active and sterile neutrinos for the case of mixing matrix $a_2$.}
\label{Fig5}
\end{figure*}

The physical reason of this difference between the different versions of the
mixing matrix is that, for the versions where the mixing matrix is either $a_1$
or $a_3$, the probability for the transition of a muon neutrino/antineutrino to
an electron neutrino/antineutrino does not contain any contributions from
sterile neutrinos, and in both cases in question these transition probabilities
are the same and are determined only by the mixing parameters of active
neutrinos. Mathematically, this is expressed by the fact that, in the analytic
expression (\ref{eq19}), only the contributions with ($i=2$, $k=1$),
($i=3$, $k=1$), and ($i=3$, $k=2$) are nonzero in the sum over $i$ and $k$ at
$\alpha=2$ and $\alpha^{\prime}=1$. All other contributions vanish identically
for mixing matrices of the $a_1$ or $a_3$ type. But in the case of a mixing
matrix belonging to the $a_2$ type, the terms with ($i=4$, $k=1$),
($i=4$, $k=2$), ($i=4$, $k=3$), ($i=5$, $k=1$), ($i=5$, $k=2$), ($i=5$, $k=3$),
and ($i=5$, $k=4$) result in nonzero contributions in the sum over $i$ and $k$,
in addition to the aforementioned nonzero contributions. At the same time, the
contributions with $i=6$ and $k<i$ again vanish identically. Nevertheless, the
mixing between all active neutrinos and two light sterile neutrinos in the case
of the mixing matrix $a_2$ provides a significant contribution to the
probabilities for the transitions of a muon neutrino/antineutrino to an
electron neutrino/antineutrino, and this contribution proves to be large owing
primarily to the term with ($i=5$, $k=4$), which has the same values for the
probabilities of the muon-neutrino transition to an electron neutrino and the
muon-antineutrino transition to an electron antineutrino. At the same time, the
difference between the probabilities for the muon-neutrino transition to an
electron neutrino and the muon-antineutrino transition to an electron
antineutrino is determined by the terms with ($i=4$, $k=1$), ($i=4$, $k=2$),
($i=4$, $k=3$), ($i=5$, $k=1$), ($i=5$, $k=2$), and ($i=5$, $k=3$), which are
different for these transitions, the terms with $k=3$ being dominant here
because of a larger value of the mass $m_3$. As can be seen from
Figs.~\ref{Fig2}--\ref{Fig5}, the values of these contributions increase as the
value of the parameter $\epsilon$ becomes larger.

Thus, a significant increase in the probabilities for the transitions of muon
neutrinos/antineutrinos to electron neutrinos/antineutrinos is determined by
the choice of the matrix $a_2$ and by the masses of light sterile neutrinos. At
the same time, this depends only slightly on the active-neutrino masses as long
as their values are substantially lower than the masses of light sterile
neutrinos and is virtually independent of the mass of the heavy sterile
neutrino. However, the largest mass among the active neutrinos ($m_3$) has a
substantial influence on the yield asymmetry, that is, on the difference in the
neutrino and antineutrino yields. Note that, in order to demonstrate the effect,
we have taken the most typical test values for the model parameters, which are
compatible with experimental data and with the results of astrophysical
observations, and verified that no qualitative changes arise in the results
upon varying these parameters, so that our result is stable in this sense.

In the following, we restrict ourselves to studying mixing matrices of just the
$a_2$ type, since only in this case the results for the probabilities of
appearance of electron neutrinos/antineutrinos may correspond to the anomaly
noticed in the LSND experiment at an energy of about $40$~MeV and a distance
of about $30$~m (see Figs.~\ref{Fig1} and \ref{Fig2}). Figure~\ref{Fig2} shows
that the probabilities for the appearance of electron neutrinos/antineutrinos
in the range between $10$ and $30$~m vary substantially, which may relate to a
negative result in searches for the LSND anomaly in the KARMEN experiment
\cite{Arm00,Arm02}, which was performed at a distance of about $17.5$~m. It can
be seen from Fig.~\ref{Fig2} that a positive result in experiments of this type
can be obtained only at high neutrino fluxes and over long experimental times,
since the transition probabilities $P(\nu_{\mu}\rightarrow\nu_e)$ and
$P(\overline{\nu}_{\mu}\rightarrow\overline{\nu}_e)$, as well as the absolute
values of their difference, are not larger than $10^{-2}$ and depend on the
parameter $\epsilon$, which should not exceed at least $0.1$.

On the left-hand panels of Fig.~\ref{Fig3}, the appearance probabilities of
electron neutrinos and antineutrinos in beams of muon neutrinos and
antineutrinos, respectively, at various coupling constants $\epsilon$ of active
and sterile neutrinos for the case of the mixing matrix $a_2$ are shown versus
the ratio of the distance $L$ from the source to the energy
$E_{\nu(\overline{\nu})}$ of neutrino/antineutrino beams at two values of the
parameter $\eta_2$, namely, $\eta_2=+\pi/6$ (upper panels) and $\eta_2=-\pi/6$
(lower panels). The graphs representing the resulting difference of the
appearance probabilities for electron neutrinos and antineutrinos (asymmetry)
are exhibited on the left-hand panels of Fig.~\ref{Fig3}. An important result,
as is seen from Fig.~\ref{Fig3}, is the change of the sign of this difference
when values of $L/E_{\nu(\overline{\nu})}\approx 2$, so that at short distances
(more precisely, distances shorter than $60$~m at the energy of
$E_{\nu(\overline{\nu})}=30$~MeV) the antineutrino yield is higher within our
model than the neutrino yield, but that at rather long distances the neutrino
yield should exceed the antineutrino yield, if $\eta_2=+\pi/6$. We emphasize
that the sign of this asymmetry changes in response to reversal of the $\eta_2$
sign (even though it is not a strictly odd function) and depends on the
parameter $L/E_{\nu(\overline{\nu})}$. Besides, the sign of the asymmetry
changes upon going over from $\delta_{CP}=-\pi/2$ to $\delta_{CP}=\pi/2$. On
the left-hand panels of Fig.~\ref{Fig3}, the results for
$P(\nu_{\mu}\rightarrow\nu_e)$ and
$P(\overline{\nu}_{\mu}\rightarrow\overline{\nu}_e)$ are shown in the same form
and versus the ratio $L/E_{\nu(\overline{\nu})}$ as it was done for the results
of the LSND and MiniBooNE experiments (see Fig.~\ref{Fig3}
in Ref.~\cite{Aguilar}). A comparison of these figures indicates that, within
the model being considered, it is possible to explain the appearance of the
accelerator anomaly in the data from those experiments.

The disappearance probability for muon neutrinos [according to Eq.~(\ref{eq19}),
it coincides with the disappearance probability for muon antineutrinos] in
beams of muon neutrinos (antineutrinos) is shown in Fig.~\ref{Fig4} as a
function of the distance from the source at various neutrino-beam energies and
various values of coupling constant $\epsilon$ of active and sterile neutrinos
for the mixing matrix $a_2$. It is seen that with the neutrino beam energy
variation the effect of sterile neutrinos to this probability behaves
qualitatively in the same fashion as in the case with the appearance probability
of electron neutrinos. That is, this probability grows monotonically as the
effect of sterile neutrinos becomes stronger.

In Fig.~\ref{Fig5}, the appearance probabilities for tau and three sterile
(left-hand panel) neutrinos and (right-hand panel) antineutrinos in beams of
muon neutrinos and antineutrinos, respectively, are shown versus the distance
from the source at the neutrino-beam energy of $30$~MeV and at the value of
$\epsilon=0.005$ of coupling constant between active and sterile neutrinos for
the mixing matrix $a_2$. One can see that the appearance of two sterile
neutrinos ($x$- and $y$-neutrinos) has the highest probability, while the
appearance probabilities for a tau active neutrino and a sterile $z$-neutrino
are relatively small. Formally, there is a difference here between the
appearance probabilities of neutrinos and antineutrinos, as this must be by
virtue of normalization conservation, on one hand, and the difference between
the appearance probabilities for electron neutrinos and antineutrinos, on the
other hand. However, the difference between the appearance probabilities for
$x$- and $y$-neutrinos and corresponding antineutrinos is relatively very small
because of the smallness of the appearance probabilities for electron neutrinos
and antineutrinos and is visually unobservable.

All these results are a feature peculiar to the above versions of the model of
active and sterile neutrinos, and they can be used in interpreting available
experimental data and in predicting the results of new relevant experiments.

\vspace{-5mm}
\section{\textnormal{Conclusion}}
\label{Section6}
Available experimental data and proposed theoretical models indicate that
neutrinos have unusual properties, and further intensive theoretical and
experimental studies are necessary to explain these properties. At the present
time, it is of greatest interest to test the existence of light sterile
neutrinos and, in the positive case, to determine their number and to fix with
a pinpoint accuracy the absolute values of both active- and sterile-neutrino
masses.

In this paper, the properties of active and sterile neutrinos versus the
sterile-neutrino masses and versus the form of the mixing matrix for active and
sterile neutrinos have been examined on the basis of a phenomenological
neutrino model with three active and three sterile neutrinos. Specifically,
this has been done by using three versions of this model, namely, the
$(3+1+2)$-, $(3+2+1)$-, and $(3+1+1+1)$-model. In this, a new parametrization
of a general mixing matrix for active and sterile neutrinos has been used,
which makes it possible to take into account the sterile-neutrino effects quite
easily. The properties of oscillations of active and sterile neutrinos in a
vacuum at test values of model parameters have been calculated. All
calculations have been performed for the case of a normal hierarchy of the mass
spectrum of active neutrinos, taking into account possible {\it CP} violation
in the lepton sector and setting the Dirac {\it CP}-phase in the
$U_{PMNS}$-matrix for active neutrinos to $-\pi/2$. The dependences of the
survival probabilities for muon neutrinos and antineutrinos and the
probabilities for their transitions to electron neutrinos and antineutrinos on
the distance from the source and on the ratio of this distance to the neutrino
energy have been determined for neutrino energies not higher than $50$~MeV.

The $(3+1+2)$, $(3+2+1)$, and $(3+1+1+1)$ versions of the general $(3+3)$-model
have been considered in order to interpret experimental neutrino oscillation
data permitting the existence of the LNSD anomaly at a neutrino energy of about
$40$~MeV and a distance of about $30$~m from the source. Graphs representing
the survival probabilities for muon neutrinos and antineutrinos and the
probabilities for their transitions to electron neutrinos and antineutrinos
have been plotted versus admissible values of the model parameters. It has been
shown (see Figs.~\ref{Fig2} and \ref{Fig3}) that this anomaly can be described
within the $(3+2+1)$-version of the model unlike the other two versions. This
version that has been found is characterized by a specific mixing matrix for
active and sterile neutrinos (namely, the matrix of the $a_2$ type), the Dirac
{\it CP} phase of $\delta_{CP}=-\pi/2$, a normal hierarchy in the mass spectrum
of active neutrinos, and a specific mass spectrum of sterile neutrinos that
includes two masses of about $1$~eV and one mass of about $7$~keV. The results
for this case on the appearance probabilities for electron neutrinos and
antineutrinos may correspond to the anomaly observed in the LSND experiment.
The model in question may also be used to describe some astrophysical data for
the cases where sterile neutrinos or warm-dark-matter particles are involved.
As for known cosmological constraints on the total number of relativistic
neutrinos \cite{Komatsy2011,Ade2013}, it should be noted that these constraints
are model-dependent. Moreover, there are the models even at the present time,
in the framework of which one can moderate substantially or even remove these
constraints, assuming different conditions of interaction and thermal
equilibrium between active and sterile neutrinos \cite{HoScherrer2013,Chu2015}.
Thus, there is the possibility of removing the possible inconsistency of the
model being considered, which contains three active and three sterile neutrinos,
with the data obtained from the cosmological observations.

It is planned to carry out in the near future a number of ground-based
experiments aimed at searches for light sterile neutrinos
\cite{Abazajian2012,Gorbunov2014,Gav2011,Bel2013,Ser2015}. For example, searches
for light sterile neutrinos have been proposed in the experiments with high
fluxes of electron antineutrinos arising in the process of $\beta$-decay of a
large number of $^{8}{\rm Li}$ nuclei \cite{Luto-Lya}. Besides, the realization
of two accelerator experiments whose results would make it possible to resolve,
with a high degree of reliability, the problem of the LSND anomaly is being
prepared. These are the OscSNS experiment in United States of America
\cite{Elnimr} and J-PARC MLF experiment in Japan \cite{Adjimura} at neutrino
energies around $40$~MeV with detectors positioned at a distance from the
neutrino source in the range from $10$ up to $100$~m. The results obtained in
the present study for the oscillation properties of active and sterile
neutrinos can be used, in particular, for interpreting the results of these and
similar experiments.

This work was supported in part by the Russian Foundation for Basic Research
(project no.\!~14-22-03040-ofi-m).

\end{document}